\documentclass[12pt]{amsart}
\usepackage{amsmath,amssymb,latexsym,graphicx}

\numberwithin{equation}{section}

\newtheorem{theorem}{Theorem}[section]

\newtheorem{remark}[theorem]{Remark}

\newtheorem*{conj}{Conjecture}
\newtheorem*{cor}{Corollary}

\begin{document}
\title[On the preservation of absolutely continuous spectrum ... ]
{On the preservation of absolutely continuous spectrum
for Schr\"odinger operators}

\author{Sergey A. Denisov}

\email{denissov@math.wisc.edu}


\date{December 22, 2004}

\address{
University of Wisconsin-Madison, Department of Mathematics,
480 Lincoln Drive, Madison WI, 53706, USA
}

\begin{abstract}
We present general principles for the preservation
of a.c. spectrum under weak perturbations. The Schr\"odinger
 operators on the strip and on the Caley tree (Bethe lattice) are considered.
\end{abstract}

\maketitle

In this paper, we consider the problem of preservation of a.c. spectrum
for Schr\"odinger operators when the perturbation (potential) is decaying
slowly at infinity.  In a recent years, lots of results in this
direction (including analysis of Jacobi matrices and Dirac operators)
were obtained \cite{Kis, DK, KS, LNS1, mnv, zlatos, Den1}. The proofs were of analytical flavor
and had roots in the approximation theory, in particular, the
theory of orthogonal polynomials of one variable \cite{Szego, Krein}. In the one-dimensional
case, most of the questions were answered and we have a rather complete
understanding of the picture. In the higher dimension, we know much less. In
a series of papers \cite{LNS2, LNS3, Den2, Den3}, authors were partially motivated by the following conjecture

\begin{conj}(B. Simon, \cite{Simon})

Consider
\[
H=-\Delta +V(x),\, x \in \Bbb R^{d}
\]
with
\[
\mathop\int\limits_{\Bbb R^{d}} \frac{V^{2}(x)}{|x|^{d-1}+1} dx < +
\infty
\]
Show that $\sigma_{ac}(H)={\Bbb R}^{+}.$
\end{conj}

\medskip
In spite of some progress in the field, this conjecture is still an open problem.

In the current paper, we consider analogous problems, but on different
domains. In the first section, we
prove one general result on preservation of absolutely continuous
spectrum for block matrices.  We also apply this result to Schr\"odinger
operator on the strip.  The second section is devoted to the discrete
Schr\"odinger operator on Caley tree (Bethe lattice) and the corresponding $L^2$
conjecture. We denote by $\mathcal{J}^p$ the standard Schatten-von
Neumann class of operators. The symbols $\mathbb {T}$ and $\mathbb{D}$  stand for the unit circle (disk) on
the complex plane. $\sigma(H)$ is the spectrum of operator $H$, $\rho(H)$-- its resolvent set. $\sigma_{ess}(H)$ and $\sigma_{ac}(H)$ denote essential and a.c. spectra, respectively. For any infinite connected tree $\Omega$, the symbol $m(\Omega)$ corresponds to the functional space of sequences decaying at infinity.

\medskip

\subsubsection* {Acknowledgment}

I am grateful to A.~Zlatos for very important remarks, to B.~Simon and T.~Spencer for useful discussions,  and to
K.~Makarov and A.~Laptev for
giving me the necessary references.
This
work was done during my stay at the Institute for Advanced Study where I
was supported by Oswald Veblen Fund.

\section{One corollary of Kato-Rosenblum theorem.}

The famous Kato-Rosenblum theorem \cite{Kato} (theorem 4.4, p. 542) says that the generalized wave operators
\mbox{$W_{\pm}(H,
H^0)$} exist for any self-adjoint bounded $H^0$  as long as
$V \in \mathcal{J}^1, V^* =V$, $H=H^0+V$.  As a corollary, $\sigma_{ac}(H^0)=
\sigma_{ac}(H)$.
In the spectral theory, it is important to study operators $H$ written in the block form.
Assume, for simplicity, that both $H_1$ and $H_2$ act in the same Hilbert space $\mathcal{H}$. Take $H$ in the following form:
\begin{equation}
H=\begin{bmatrix}
H_1&V\\
V^*&H_2\\
\end{bmatrix} \label{hamil}
\end{equation}
An interesting problem is to consider the off-diagonal operator
$$
\begin{bmatrix}
0&V\\
V^*&0\\
\end{bmatrix}
$$
as a perturbation of
$$
H^0=\begin{bmatrix}
H_1&0\\
0&H_2\\
\end{bmatrix}
$$
and develop the corresponding perturbation theory.
We will need to use the following result contained in \cite{Mak}

\begin{theorem}\cite{Mak}. \label{n0}
If $H_{1(2)}$ are selfadjoint bounded operators, $\sigma(H_1)\subset (a_1, a_2)\subset \rho(H_2)$  and $V\in {\mathcal{J}}^2$, then
$\sigma_{ac}(H)=\sigma_{ac}(H^0)$, with $H$ given by (\ref{hamil}).
\end{theorem}

Indeed, by Kato-Rosenblum theorem, a.c. part of the spectrum is invariant under the finite-rank perturbations. Therefore, we can assume $\|V\|$ is as small as we wish. Then, due to the separation of $\sigma(H_1)$ and $\sigma(H_2)$, one can show the existence of reducing graph subspaces for $H$ of the form $(x, \Gamma_1 x)$ and $(\Gamma_2 x, x)$, with some operators $\Gamma_{1(2)}$ (theorem 7.6 of \cite{Mak}). Then, using $\Gamma_{1(2)}$, one can see (theorem 5.5,\cite{Mak}) that $H$ is unitarily equivalent to the operator

$$
L=\begin{bmatrix}
H_1+Q_1&0\\
0&H_2+Q_2\\
\end{bmatrix}
$$
where $Q_{1(2)}\in{ \mathcal{J}}^1$. Application of the Kato-Rosenblum theorem again concludes the argument.

We will prove the following statement

\begin{theorem}\label{n1}
Let $H_1, H_2$ be two bounded self-adjoint operators in the Hilbert space
$\mathcal{H}$. Assume that $\sigma_{ess}(H_2)
\subseteq [b, + \infty]$ and $[a,b]
\subseteq \sigma_{ac}(H_1), (a <b)$.  Then, for any $V \in {\mathcal{J}}^2$,  we have that $[a,b] \subseteq \sigma_{ac}(H)$, with $H$ given by (\ref{hamil}).
\end{theorem}

\begin{remark}   The case when
the spectra of $H_1$ and $H_2$ overlap is more involved as we will see, but the phenomena
is more or less the same:  the influence of $H_2$ on $[a,b]$ is small
and the Hilbert-Schmidt off-diagonal perturbation does not change
a.c. spectrum of $H$ on $[a,b]$.
\end{remark}

\begin{proof}
Take any $\epsilon >0$ and let $b_\epsilon=b-\epsilon$. Let us show that $[a,b_\epsilon]
\subseteq \sigma_{ac}(H)$.  Since the finite-rank perturbations do not
change a.c. spectrum and $\sigma_{ess}(H_2)
\subseteq [b, + \infty)$, we can always assume that
\[
\sigma(H_2) \subseteq
[b-\epsilon/2, + \infty)
\]

Denote $R^{(1)}_z=(H_1 -z)^{-1},
R^{(2)}_z=(H_2 - z)^{-1}$.

We can write
$$
R_z=(H-z)^{-1}=\begin{bmatrix}
A_{11}(z)&A_{12}(z)\\
A_{21}(z)&A_{22}(z)\\
\end{bmatrix}
$$
If $x=A_{11}(z)f+A_{12}(z)g, y=A_{21}(z)f+A_{22}(z)g$, then
\begin{equation}
\begin{cases}
(H_{1}-z)x+Vy=f\\
V^*x+(H_{2}-z)y=g
\end{cases}
\begin{cases}
x=R^{(1)}_z f-  R^{(1)}_z VR^{(2)}_zg+R^{(1)}_z VR^{(2)}_z V^*x\\
y=R^{(2)}_z g - R^{(2)}_z V^* R^{(1)}_z f + R^{(2)}_z V^* R^{(1)}_z V y
\end{cases}
\label{st1}
\end{equation}

Consider the following operator
$M=(H-b_\epsilon)^2 P_{b_\epsilon}$, where $P_{b_\epsilon}=P_{(-\infty, b_\epsilon]}$-- spectral projector for $H$.
We have the following general formula
\begin{equation}
M=\frac{1}{2\pi i} \int\limits_\gamma (z-b_\epsilon)^2 R_z dz \label{mmm}
\end{equation}
where $\gamma$ is a contour, say, rectangle, that crosses the real axes at the point $b_\epsilon$ with its right side
and ``encircles" the parts of $\sigma(H_1)$ and $\sigma(H)$ below the point $b_\epsilon$. The integration is anticlockwise.

Using (\ref{mmm}) and the formulas (\ref{st1}) for $x$ and $y$, we have
\begin{equation}
M=M_0+T=\begin{bmatrix}
(H_1-b_\epsilon)^2 P^{(1)}_{b_\epsilon} & S\\
S^* & 0\\
\end{bmatrix}+
\begin{bmatrix}
T_{11} & T_{12}\\
T^*_{12} & T_{22}\\
\end{bmatrix} \label{repr1}
\end{equation}
where
\[
S=-\frac{1}{2\pi i} \int\limits_\gamma (z-b_\epsilon)^2 R_z^{(1)} V R_z^{(2)} dz
\]
\[
T_{11}=
\frac{1}{2\pi i} \int\limits_{\gamma}
(z-b_\epsilon)^2 R_z^{(1)} V R_z^{(2)} V^* A_{11}(z) dz
\]
\[
T_{12}=
\frac{1}{2\pi i} \int\limits_{\gamma}
(z-b_\epsilon)^2 R_z^{(1)} V R_z^{(2)} V^* A_{12}(z) dz
\]
\[
T_{22}=
\frac{1}{2\pi i} \int\limits_{\gamma}
(z-b_\epsilon)^2 R_z^{(2)} V^* R_z^{(1)} V A_{22}(z) dz
\]
Due to the regularization factor $(z-b_\epsilon)^2$, the operator $T \in{ \mathcal{J}}^1$.
Therefore, by Kato-Rosenblum, $
\sigma_{ac} (M)=
\sigma_{ac} (M_0)$. For $S$, we have an expression
\begin{equation}
S=-\frac{1}{2\pi i} \int\limits_\gamma
(z-b_\epsilon)^2 \int\limits \frac{dP^{(1)}_\lambda}{\lambda-z} V \int\limits_{b-\epsilon/2}^\infty \frac{dP^{(2)}_t}{t-z}=
\int\limits_{-\infty}^{b_\epsilon} (\lambda-b_\epsilon)^2 dP^{(1)}_\lambda V R_\lambda^{(2)} \label{ss}
\end{equation}

Now, we are going to use theorem \ref{n0} in the following way. Consider
the operator

$$
\hat{H}=\begin{bmatrix}
\hat{H}_1 &V\\
V^*&H_2\\
\end{bmatrix}
$$
where
$\hat{H}_1=H_1P^{(1)}_{b_\epsilon}+(b_\epsilon+\epsilon/4)P^{(1)}_{(b_\epsilon, \infty)}$.
For $\hat {H}$, theorem \ref{n0} is now applicable because the spectra are separated and we therefore have
$[a,b_\epsilon]\subseteq\sigma_{ac}(\hat{H})$. Thus, for the corresponding
$\hat{M}=(\hat{H}-b_\epsilon)^2 \hat{P}_{b_\epsilon}$, we get
$(0,(b_\epsilon-a)^2)\subseteq \sigma_{ac} (\hat{M})$ by the Spectral theorem. On the other hand, $\hat{M}$ allows
the representation similar to (\ref{repr1}):
\[
\hat{M}=\hat{M}_0+\hat{T}
\]
From the choice of $\hat{H}_1$ and the formula (\ref{ss}) for  $S$, we have $\hat{M}_0=M_0$. Since
$\hat{T}$ is trace class, we obtain $\sigma_{ac} (M)=\sigma_{ac} (\hat{M})$. Therefore,
$(0,(b_\epsilon-a)^2)\subseteq \sigma_{ac} (M)$ and by the Spectral theorem $[a,b_\epsilon]\subseteq \sigma_{ac} (H)$.
\end{proof}

It is easy to generalize this result to the case when $V$ is a relative Hilbert-Schmidt perturbation and apply it  to continuous Schr\"odinger operator on the
strip $\Pi=\{0<x<\infty, 0<y<\pi\}$.  Let
$$
L=-\Delta +Q(x,y)
$$
and impose Dirichlet conditions on the boundary of $\Pi$. Then
\begin{cor} If $\mathop{sup}\limits_{0\leq y \leq \pi}|Q(x,y)|\in L^2
(\Bbb R^+) \cap  L^\infty (\Bbb{R}^+)$, then $[1,4] \subseteq \sigma_{ac}(H)$.
\end{cor}
This result follows from the matrix
representation of $L$.  For $f(x,y)\in L^2(\Pi)$,

$$
f(x,y)=\sqrt{\frac{2}{\pi}}\mathop\sum\limits^{\infty}_{n=1} \sin(ny)
f_n (x),  f_n(x)=\sqrt{\frac{2}{\pi}}\mathop\int\limits_{0}^{\pi}
f(x,y) \sin(ny)dy
$$
and the matrix representation of $L$ is as follows
\begin{equation}
L=\left[ \begin{array}{ccc}
\displaystyle -\frac{d^2}{dx^2} +Q_{11}(x)+1_{\, |D} & Q_{12}(x) & \ldots\\
Q_{21}(x) & \displaystyle -\frac{d^2}{dx^2}+Q_{22}(x)+4_{\, |D} & \ldots \\
\ldots & \ldots & \ldots \\
\end{array} \right] \label{matr}
\end{equation}
$$
Q_{lj}(x)=\frac{2}{\pi}\mathop\int\limits_0^{\pi} Q(x,y)\sin(ly)\sin(jy)dy
$$
Since $Q_{11}(x)\in L^2(\Bbb R^+)$ we can use \cite{DK} and theorem \ref{n1} to conclude
$[1,4]\subseteq \sigma_{ac}(H)$.
Indeed, one can proceed in the following way.
Consider the  operator
\[
Z=\begin{bmatrix}
A_1 & B\\
B^* & A_2\\
\end{bmatrix}
\]
where $A_2$ is the matrix of $L$ with the first row and column deleted.
In theorem~\ref{n1}, $H_{1(2)}$  act in the same space. In our case, we can always extend the matrix trivially by, say, identity and take
\[
A_1=
\left[ \begin{array}{cccc}
\displaystyle -\frac{d^2}{dx^2} +Q_{11}(x)+1_{\, |D} & 0 & 0 &\ldots\\
0 & 1 & 0 & \ldots\\
0 & 0 & 1 & \ldots\\
\ldots & \ldots & \ldots & \ldots \\
\end{array} \right]
\]
where $A_1$ acts now in the same space $\mathcal{H}=\prod\limits_{n=2}^\infty L^2(\mathbb{R}^+)$ as $A_2$.
Obviously,
\[
B=
\left[ \begin{array}{ccc}
Q_{12}(x) & Q_{13}(x) &  \ldots\\
0 & 0 &  \ldots\\
\ldots & \ldots  & \ldots \\
\end{array} \right]
\]
Without the loss of generality, we can always assume that $\|Q\|$ is small.
Taking the inverse of the matrix $Z$ and using the formula for inversion of the block operators

\[
Z^{-1}=\begin{bmatrix}
(A_1-BA_2^{-1}B^*)^{-1} & -A_1^{-1}B (A_2- B^*A_1^{-1} B)^{-1}\\
-A_2^{-1} B^*(A_1- BA_2^{-1}B^*)^{-1}  & (A_2- B^*A_1^{-1} B)^{-1}\\
\end{bmatrix}
\]
we can apply the theorem \ref{n1} to $Z^{-1}$ to get the corollary immediately.

In \cite{mnv1} (see also
\cite{ls,lw}), the case of matrix-valued potential was considered.
Using this technique, it is conceivable (and authors of
\cite{mnv1} intend to do that) that $L^2$ conjecture can be proved
for Schr\"odinger operators with matrix-valued potential of the
form $V=D+v$, where $D$-- diagonal and $\|v\|\in
L^2(\mathbb{R}^+)$. In our case, that would allow to conclude that
$\sigma_{ac}(L)=[1,\infty)$.

\section{$L^2$ spectral conjecture for Schr\"odinger operator on
the Caley tree (Bethe lattice).}

In this section, we consider the Caley tree (Bethe lattice) $\mathbb{B}$ and the
discrete Laplacian on it
$$
(H_0 u)_n =\mathop\sum\limits_{|i-n|=1} u_i
$$
We remind that the Caley tree is an infinite graph with no closed loops and a fixed degree (number of nearest neighbors)
at each vertex. We will denote by $|x-y|$ the length of the shortest path connecting $x$ and $y$.
Assume, for simplicity, that the degree at each point is equal to $3$. All arguments can be easily adjusted to
the general case and even to some modifications of the Laplacian.
It is both known and easy to see that $\sigma(H_0)=[-2\sqrt{2}, 2 \sqrt{2}]$ and the
spectrum is purely absolutely continuous. Let $H=H_0+V$, where $V$ is a potential. The natural question to ask is
under which conditions on $V$ the a.c. spectrum of $H_0$ is preserved. The Bethe lattice has a great advantage compared to $\mathbb{Z}^{d}$: there is only one path connecting any two points. That usually makes it possible to set up a recursion relation to control different quantities (see, e.g., the proof of delocalization for the Anderson model \cite{klein, aiss}). The result below and its proof has their roots in the theory of orthogonal polynomials on the unit circle \cite{khr} and on the real line. One could call it the multidimensional step-by-step sum rule (using the term recently introduced into the subject \cite{KS, simon1}). At the same time, the idea has a clear physical meaning, as will be discussed later on. We hope that it might give some insight to the solution (positive or negative) of $L^2$ conjecture on $\mathbb{Z}^d$. Scattering on the special graphs and networks was studied before (see, e.g., \cite{pavlov, froese}).

Consider any vertex $O$. It is connected to its neighbors by three edges. Delete one edge together with the corresponding part of the tree stemming from it. What is left will be called $\mathbb{B}_{O}$.
The degree of $O$ within $\mathbb{B}_{O}$ is equal to $2$.   The main result of this section is the following

\begin{theorem}\label{n2}
If $V\in \ell^\infty(\mathbb{B})\cap m(\mathbb{B}_O)$ and
\begin{equation}
\mathop\sum\limits_{n=0}^{\infty}
\frac{1}{2^{n}} \mathop\sum\limits_{x\in \mathbb{B}_{O}, |x-O|=n} V^2(x)
< +\infty \label{ell2}
\end{equation}
then

$$
[-2\sqrt{2}, 2
\sqrt{2}]=\sigma_{ac}(H_{|\mathbb{B}_{O}})\subseteq \sigma_{ac}(H)
$$
\end{theorem}

\begin{proof}   The potential is bounded, therefore $H$ is well defined as the self-adjoint bounded operator.
Since $\mathbb{B}_{O}$ is connected to the other part of
$\mathbb{B}$ by just one edge, the rank-two perturbation argument \cite{Kato} (theorem 4.4, p. 542)
allows us to focus on $H_{|\mathbb{B}_{O}}$ only, i.e. we only need to
show that $\sigma_{ac}(H_{|\mathbb{B}_{O}})=[-2\sqrt{2}, 2\sqrt{2}]$.
Because $V\in m(\mathbb{B}_{O})$
\footnote{That condition can probably be discarded. We impose it for simplicity.}
, the Weyl theorem (theorem 5.35, p. 244, \cite{Kato}) yields $\sigma_{ess} (H_{|\mathbb{B}_{O}})=[-2\sqrt{2}, 2\sqrt{2}]$.
Let $\delta_{X}$ be the discrete delta-function with support at the vertex $X$.
We will show that the spectral measure corresponding to $\delta_{O}$ has a.c. component filling $[-2\sqrt{2}, 2\sqrt{2}]$. That would finish the proof.

First, assume that $V_{|\mathbb{B}_{O}}$ has a finite support. We introduce

$$
m(\lambda)=((H_{|\mathbb{B}_{O}}-\lambda)^{-1} \delta_{O}, \delta_{O})
$$
$$
m_1(\lambda)=((H_{|\mathbb{B}_1} -\lambda)^{-1} \delta_{O_{1}}, \delta_{O_{1}})
$$
$$
m_2(\lambda)=((H_{|\mathbb{B}_2}-\lambda)^{-1} \delta_{O_{2}}, \delta_{O_{2}})
$$
where $O_{1(2)}$ - neighbors of $O$ in $\mathbb{B}_{O}$,
$H_{|\mathbb{B}_j}$ - restrictions of $H_{|\mathbb{B}_{O}}$ to the branches $\mathbb{B}_j$ having
``roots'' at $O_j, j=1,2$.
Functions $m, m_1, m_2$ have positive imaginary parts in $\mathbb{C}^+$.
The following is straightforward

\[
-m_{1}m - m_2 m=-V(O)m+\lambda m+1
\]
and
\begin{equation}
m(\lambda)=- \frac{1}{m_{1}(\lambda)+m_{2}(\lambda)-V(O)+\lambda} \label{recur}
\end{equation}
Recurrence relation (\ref{recur}) has the fundamental importance.
If $V \equiv 0$, we have

$$
m_0(\lambda)=- \frac{1}{2m_0(\lambda)+\lambda}, \quad
m_0(\lambda)= \frac{-\lambda +\sqrt{\lambda^{2}-8}}{4}
$$
(\ref{recur}) can be also used to obtain an asymptotical
expansion of $m(\lambda)$ as $\lambda \rightarrow \infty$.
We will need only a couple of coefficients in this expansion.
Clearly, $m_j(\lambda)=-\lambda^{-1}+O(\lambda^{-2}),
j=1,2$ and

$$
m(\lambda)=-\frac{1}{-V(O)+\lambda-2\lambda^{-1}+O(\lambda^{-2})}=
$$

$$
=-\frac{1}{\lambda[1-(V(O)\lambda^{-1}+2\lambda^{-2}+O(\lambda^{-3})]}=
$$

$$
=-\lambda^{-1}[1+V(O)\lambda^{-1}
+2\lambda^{-2}+V^2(O)\lambda^{-2}+O(\lambda^{-3})]=
$$

$$
=-[\lambda^{-1}+V(O)\lambda^{-2}+(2+V^{2}(O))\lambda^{-3}+O(\lambda^{-4})]
$$
One can iterate (\ref{recur}) to get an interesting analog of the continued fraction.
Knowing $m_0(\lambda)$ and recursion relation, we conclude that
$m(\lambda)$  is  meromorphic  in   $\Bbb C \backslash \{[-2\sqrt{2}, 2
\sqrt{2}]\}$. Introduce $M(\lambda)=m_1 +m_2-V(O)+\lambda$. Then

\begin{equation}
m(\lambda)=-\frac{1}{M(\lambda)} \label{inverse}
\end{equation}

 In the next paragraph, we repeat the calculation done by B. Simon \cite{simon1} who was focusing on the Jacobi matrices.
Let us map $\mathbb{C}\setminus\{[-2\sqrt2, 2\sqrt2]\}$ to the unit disk $\mathbb{D}$ by considering
$f(z)=-m(\sqrt{2}(z+z^{-1}))$
and
$F(z)=-M(\sqrt{2}(z+z^{-1}))$.
Notice that in the free case $f_0(z)=z/\sqrt{2}$.
Let $z_k^+$ and $p_k^+$ be positive zeroes/poles of $f(z)$, respectively. Denote the negative
zeroes/poles by $z_k^-$ and $p_k^-$.
We order them as follows
\[
0<z_1^+ <z_2^+ <\ldots<1, 0<p_1^+ <p_2^+ <\ldots<1,
\]
\[
0>z_1^- >z_2^- >\ldots>-1, 0>p_1^- >p_2^- >\ldots>-1
\]

We have
\begin{equation}
f(z)=-\frac{1}{F(z)}, \qquad \frac{\Im f(z)}{\Im F(z)}=|f(z)| ^2 \label{star}
\end{equation}
The function $f(z)$ has positive/negative imaginary parts in $\mathbb{C}^{\pm}\cap \mathbb{D}$.
Then, by Simon's theorem \cite{simon1}, we have a  multiplicative representation for $f(z)$

\begin{equation}
f(z) =zB(z)\exp\Big(\frac{1}{2\pi}\mathop\int\limits_{-\pi}^{\pi}
\frac{t+z}{t-z} \ln |f(t)|d
\theta\Big), t=e^{i\theta} \label{multipl}
\end{equation}

$$
B(z)=B_1(z)B_2^{-1}(z)
$$

$$
B_1(z)=\prod \limits_{k} \frac{|z_{k}|}{z_{k}} \cdot
\frac{z_{k}-z}{1-z_{k}z},\,
B_2(z)=\prod \limits_{k} \frac{|p_{k}|}{p_k}\cdot
\frac{p_{k}-z}{1-p_{k}z}
$$
Let us divide the both sides of (\ref{multipl}) by $z$, take $\ln$, and expand into
the Taylor series around $0$. We have

$$
f(z)=\frac{1}{\sqrt{2}}
\frac{z}{z^{2}+1}+
\frac{V(O)}{2}\cdot
\frac{z^{2}}{(z^{2}+1)^2}+
\frac{2+V^{2}(O)}{2\sqrt{2}} \cdot
\frac{z^{3}}{(z^{2}+1)^3}+O(z^4)
$$

$$
=\frac{1}{\sqrt 2}z+\frac{V(O)}{2}z^2 + \frac{V^{2}(O)}{2 \sqrt
2}z^3+O(z^4);
$$

$$
\ln\Big(\frac{f(z)}{z}\Big)=-\frac{1}{2} \ln 2+\frac{V(O)}{\sqrt
2}z+\frac{V^2(O)}{4} z^2 + \ldots;
$$

$$
\frac{1}{2 \pi}\mathop\int\limits_{-\pi}^{\pi} \frac{t+z}{t-z}
\ln|f(t)|d \theta=\frac{1}{2\pi} \mathop\int\limits_{-\pi}^{\pi}
\ln|f(t)|(1+2zt^{-1}+2z^{2}t^{-2}+\ldots) d\theta;
$$

$$
\ln
\Big(\frac{z_{k}^+-z}{1-z_{k}^+ z}\Big)=\ln[(z_{k}^+ -z)(1+z_{k}^+ z+{z_{k}^+}^2
z^2 + \ldots )]=
$$
$$
=\ln z_k^++z(z_k^+ - {z_{k}^+}^{-1})+z^2({z_{k}^+}^{2}-{z_{k}^+}^{-2}) / 2 +O(z^3)
$$
Comparing the first and the third coefficients, we obtain
\begin{equation}
- \frac{1}{2} \ln 2=\frac{1}{2 \pi}
  \mathop\int\limits_{-\pi}^{\pi} \ln |f(t)|d \theta +
\mathop\sum\limits_{k} \ln|z_k|-\mathop\sum\limits_{k} \ln|p_k| \label{eq1}
\end{equation}
and
\begin{equation}
\frac{V^2(O)}{4}=\frac{1}{\pi}
\mathop\int\limits_{-\pi}^{\pi} \ln|f(t)|t^{-2}d \theta
+\mathop\sum\limits_{k} \frac{|z_k|^2-|z_k|^{-2}}{2} -
\mathop\sum\limits_{k} \frac{|p_k|^2 - |p_k|^{-2}}{2} \label{eq2}
\end{equation}
Take (\ref{eq1})-$1/2\, \cdot$ (\ref{eq2}). We then have

$$
-\frac{1}{2} \ln 2-\frac{V^2(O)}{8}=\frac{1}{2\pi}
\mathop\int\limits_{-\pi}^{\pi} \ln|f(t)|(1-t^{-2})d \theta -
\Big\{\mathop\sum\limits_{k} Y(\widetilde E_k)-
\mathop\sum\limits_k Y(E_k)\Big\}
$$
where

$$
Y(-E)=Y(E)=\frac{|z|^2-|z|^{-2}}{4}-\ln|z|,
 E=\sqrt 2 (|z| + |z|^{-1}), |z|<1, E>2\sqrt{2}
$$
and
\[
E_k^{\pm}=\pm \sqrt{2} (|p_k^{\pm}|+|p_k^{\pm}|^{-1}),
\widetilde{E}_k^{\pm}=\pm \sqrt{2} (|z_k^{\pm}|+|z_k^{\pm}|^{-1})
\]

Recall (\ref{star})
$$
\frac{1}{2\pi} \mathop\int\limits_{0}^{\pi} \frac{\ln\  \Im
f}{\ln\Im F} (1-\cos 2 \theta)d\theta=
$$
$$
=-\frac{1}{2}
\ln 2-\frac{V^{2}(O)}{8}+\mathop\sum\limits_k Y(\widetilde
E_k^+)-\mathop\sum\limits_k Y(E_k^+)+\mathop\sum\limits_k Y(\widetilde
E_k^-)-\mathop\sum\limits_k Y(E_k^-)
$$

For the spectral measures,
$$
\Im  m(\lambda+i0)=\pi \sigma_O'(\lambda), \quad \lambda \in [-2\sqrt
2, 2 \sqrt 2]
$$

$$
\Im M(\lambda+i0)=\Im  m_1(\lambda+i0)+\Im
 m_2(\lambda+i0)=\pi (\sigma'_{O_1}(\lambda) +\sigma'_{O_2} (\lambda))
$$
Then, the multidimensional step-by-step sum rule is

$$
\frac{1}{8\pi} \mathop\int\limits_{-2\sqrt 2}^{2\sqrt 2}
\sqrt{8-\lambda^{2}}
\ln
[\pi \sigma'_O(\lambda)] d\lambda=\frac{1}{8\pi}
\mathop\int\limits_{-2\sqrt2}^{2\sqrt2} \sqrt{8-\lambda^{2}} \ln \Big(\pi \frac{\sigma'_{O_1}(\lambda)
+\sigma'_{O_2}(\lambda)}{2}\Big) d\lambda
$$

$$
-\frac{V^2(O)}{8}+\mathop\sum\limits_k Y(\widetilde
E_k)-\mathop\sum\limits_k Y(E_k)
$$

Let us show that
\begin{equation}
\sum\limits_{k} Y(\widetilde E_k) \geq \sum\limits_k Y(E_k) \label{monoton}
\end{equation}

Indeed, $Y(2\sqrt{2})=0$, $Y(E)$ is even, negative, and decreasing in $E>2\sqrt{2}$.
Notice that $E_k$ are poles of $m(\lambda)$, $\widetilde{E}_k$ are zeroes of $m(\lambda)$ (and poles of $M(\lambda)$).
From (\ref{inverse}) and integral representation of $m(\lambda)$ we have the interlacing property

$$
E_1^- < \widetilde{E}_1^-< E_2^-< \ldots <-2\sqrt{2}
$$
$$
2\sqrt{2}<\ldots<E_2^+< \widetilde{E}_1^+<E_1^+
$$
So, (\ref{monoton}) holds and

\begin{equation}
\frac{1}{\pi} \mathop\int\limits_{-2\sqrt 2}^{2\sqrt 2}
\sqrt{8-\lambda^{2}}
\ln
[\pi \sigma'_O(\lambda)] d\lambda \geq \frac{1}{\pi}
\mathop\int\limits_{-2\sqrt 2}^{2\sqrt2} \sqrt{8-\lambda^{2}} \ln
\Big(\pi \frac{\sigma'_{O_1}(\lambda) + \sigma'_{O_2}(\lambda)}{2}\Big)
d\lambda-V^2(O) \label{thepoint}
\end{equation}
We have
$$
\frac{1}{\pi} \mathop\int\limits_{-2\sqrt2}^{2\sqrt2} \sqrt{8-\lambda^{2}} \ln
[\pi \sigma'_O(\lambda)] d\lambda\geq
$$
$$
\geq \frac{1}{2} \Big[ \frac{1}{\pi}
\mathop\int\limits_{-2\sqrt2}^{2\sqrt2} \sqrt{8-\lambda^{2}} \ln[\pi \sigma'_{O_1}(\lambda)]
d\lambda + \frac{1}{\pi}
\mathop\int\limits_{-2\sqrt2}^{2\sqrt2} \sqrt{8-\lambda^{2}}\ln[\pi \sigma'_{O_2}(\lambda)]
d\lambda \Big] -V^2(O)
$$
Iterate this estimate till we leave the support of $V$.
Then, using the precise expression for $m_0(\lambda)$ and $\sigma'_0(\lambda)=\pi^{-1}\Im m_0(\lambda)$, one has

$$
\frac{1}{\pi}\mathop\int\limits_{-2\sqrt2}^{2\sqrt2} \sqrt{8-\lambda^{2}} \ln
\Big(\frac{4 \sigma'(\lambda)}{\sqrt{8-\lambda^{2}}}\Big) d\lambda
\geq -\mathop\sum\limits_{n=0}^{\infty} \frac{1}{2^{n}}
\mathop\sum\limits_{|j-O|=n} V^{2}(j)
$$
It is instructive to introduce the notion of relative entropy \cite{KS, simon1}
\[
S(\mu|\nu)=
-\int \ln\left( \frac{d\mu}{d\nu}\right) d\mu
\]
and write
\begin{equation}
S(\sigma_0 | \sigma) \geq - \frac{1}{4}
\mathop\sum\limits_{n=0}^{\infty} \frac{1}{2^{n}}
\mathop\sum\limits_{|j-O|=n} V^2(j) \label{bound}
\end{equation}
The bound (\ref{bound}) can be extended to all potentials satisfying (\ref{ell2}).
This is a standard argument \cite{KS} that uses semicontinuity of the entropy and the fact that
$d\sigma_O^{(k)}$ converges weakly to $d\sigma_O$ as $k\to\infty$.
Here $d\sigma_O^{(k)}$ corresponds to the potential truncated to the ball in $\mathbb{B}_{O}$  of radius $k$ centered at $O$. The estimate (\ref{bound}) implies that the spectral measure for the delta function centered at the origin has an a.c. part supported by $[-2\sqrt2, 2\sqrt 2]$.

\end{proof}

Unlike in the case of Jacobi matrices, we are unable to prove the real sum rule, i.e., the criteria for the perturbation to belong to a certain class (see, e.g., \cite{KS}). Indeed, we essentially use the Jensen inequality. Even more, we provided no information on the discrete spectrum. It is an interesing problem to obtain an adequate analog of the Lieb-Thirring inequality. 

In (\ref{thepoint}), we could have
used

$$
\frac{1}{\pi} \mathop\int\limits_{-2\sqrt2}^{2\sqrt2} \sqrt{8-\lambda^{2}}
\ln[\pi
\sigma_{O}'(\lambda)]  d\lambda \geq \frac{1}{\pi}
\mathop\int\limits_{-2\sqrt2}^{2\sqrt2} \sqrt{8-\lambda^{2}} \ln
\Big(\pi \frac{\sigma'_{O_1}(\lambda)}{2}\Big) d\lambda -V^2(O)
$$
This inequality has a perfect physical meaning. If the branches $\mathbb{B}_1$ and $\mathbb{B}_2$ are not connected
and $H_{|\mathbb{B}_1}$ is such that a particle can scatter from $O_1$, then it can
scatter from $O$ also, regardless of what happens at the other
branch $\mathbb{B}_2$. This is a general fact, which can be adjusted to any tree and to different modifications of the corresponding Laplacian.  In particular, it is true that under the conditions of
theorem {\ref{n2}},  the a.c. component of the spectral measure corresponding to {\it any} vertex in $\mathbb{B}$ is supported by $[-2\sqrt 2, 2\sqrt 2]$, i.e. the scattering is possible
from { \it any} vertex of the lattice.

For Caley trees, the right quantity to study is an entropy
$S(\sigma_0 |\sigma)$ and an accurate bound is (\ref{bound}). The method seems to be adjustable to the decorated trees, for which there are many paths connecting two points.

For the general lattice, using the rank-one perturbation technique, one can compute how the ``entropy at the site" changes upon making the potential $0$ at that site. This simple calculation reveals some analytic structure. It certainly says nothing about the global scattering picture, for this argument does not feel the lattice itself. We do not discuss these issues here.
The proof of theorem \ref{n2} is rather easy, but it is very different from what was used before to deal with multidimensional scattering problems. We believe it might be of some interest.

The theorem {\ref{n2}} is optimal in a sense. One can easily find \cite{Kis} the spherically symmetric potential $V(x)=V(|x|)\in \ell ^p (\mathbb{Z}^+), p>2$, for which the spectral measure corresponding to the origin $O$ has no a.c. component. That can be done using the reduction to the one-dimensional discrete Schr\"odinger operator.

\end{document}